\pdfoutput=1

\PassOptionsToPackage{table,svgnames}{xcolor}

\documentclass[11pt,fleqn]{article}

\usepackage[preprint]{acl}

\usepackage{times}
\usepackage{latexsym}

\usepackage[T1]{fontenc}

\usepackage[utf8]{inputenc}

\usepackage{microtype}

\usepackage{inconsolata}

\usepackage{graphicx}

\usepackage{multirow} 
\usepackage{booktabs}
\usepackage[most]{tcolorbox}
\usepackage{enumitem}
\usepackage{listings}

\usepackage{colortbl}

\usepackage[export]{adjustbox}

\title{SPAR: Scholar Paper Retrieval with LLM-based Agents for Enhanced Academic Search}

\author{
    Xiaofeng Shi\textsuperscript{1}\thanks{Equal contribution.}\thanks{Corresponding author. Email: \texttt{xfshi@baai.ac.cn}} \quad
    Yuduo Li\textsuperscript{1,2}\footnotemark[1]\thanks{Work done during internship at BAAI.} \quad
    Qian Kou\textsuperscript{1}\footnotemark[1] \quad
    Longbin Yu\textsuperscript{1} \quad
    Jinxin Xie\textsuperscript{1} \quad
    Hua Zhou\textsuperscript{1}\thanks{Project leader.} \\
    \textsuperscript{1}Beijing Academy of Artificial Intelligence (BAAI) \\
    \textsuperscript{2}Beijing Jiaotong University (BJTU)
}

\begin{document}
\maketitle

\begin{abstract}
Recent advances in large language models (LLMs) have opened new opportunities for academic literature retrieval. However, existing systems often rely on rigid pipelines and exhibit limited reasoning capabilities. We introduce \textbf{SPAR}, a multi-agent framework that incorporates \textit{RefChain-based query decomposition} and \textit{query evolution} to enable more flexible and effective search. To facilitate systematic evaluation, we also construct \textbf{SPARBench}, a challenging benchmark with expert-annotated relevance labels. Experimental results demonstrate that SPAR substantially outperforms strong baselines, achieving up to \textbf{+56\%} F1 on AutoScholar and \textbf{+23\%} F1 on SPARBench over the best-performing baseline. Together, SPAR and SPARBench provide a scalable, interpretable, and high-performing foundation for advancing research in scholarly retrieval. Code and data will be available at: \url{https://github.com/xiaofengShi/SPAR}
\end{abstract}

\section{Introduction}

Effective academic paper retrieval is fundamental to research. As scientific literature continues to grow exponentially, researchers are increasingly challenged by the need to locate not just superficially relevant papers, but comprehensive and interconnected works that span multiple subtopics, time periods, and academic communities~\cite{gusenbauer2020academic}. While traditional academic search engines such as Google Scholar~\cite{vine2006google} support basic keyword queries well, they often fall short in supporting complex, multi-intent queries that require deeper contextual understanding or reference-based exploration.

Consider the query: \emph{“Show some cutting-edge technological advancements on how to improve the generalization ability of machine learning models across multiple domains.”} This query implicitly demands up-to-date results, an understanding of “generalization” in a machine learning context, and coverage across multiple subfields. Existing systems tend to either return overly generic results or fail to capture the full semantic scope of such queries, leading to time-consuming manual filtering by the user.

Recent advances in large language models (LLMs)~\cite{achiam2023gpt,team2023gemini,liu2024deepseek,yang2025qwen3} have enabled promising developments in information retrieval, including query rewriting, document retrieval, and ranking~\cite{zhu2023large}. In the academic domain, these capabilities offer potential to support more intelligent, context-aware search experiences. However, academic research involves more than retrieving documents matching a user query: researchers often explore citation networks, follow references recursively, and synthesize insights across multiple papers. These behaviors, central to scholarly discovery, remain underexplored in current LLM-based retrieval systems.

\begin{figure}[htp]
  \centering
  \includegraphics[width=0.8\linewidth]{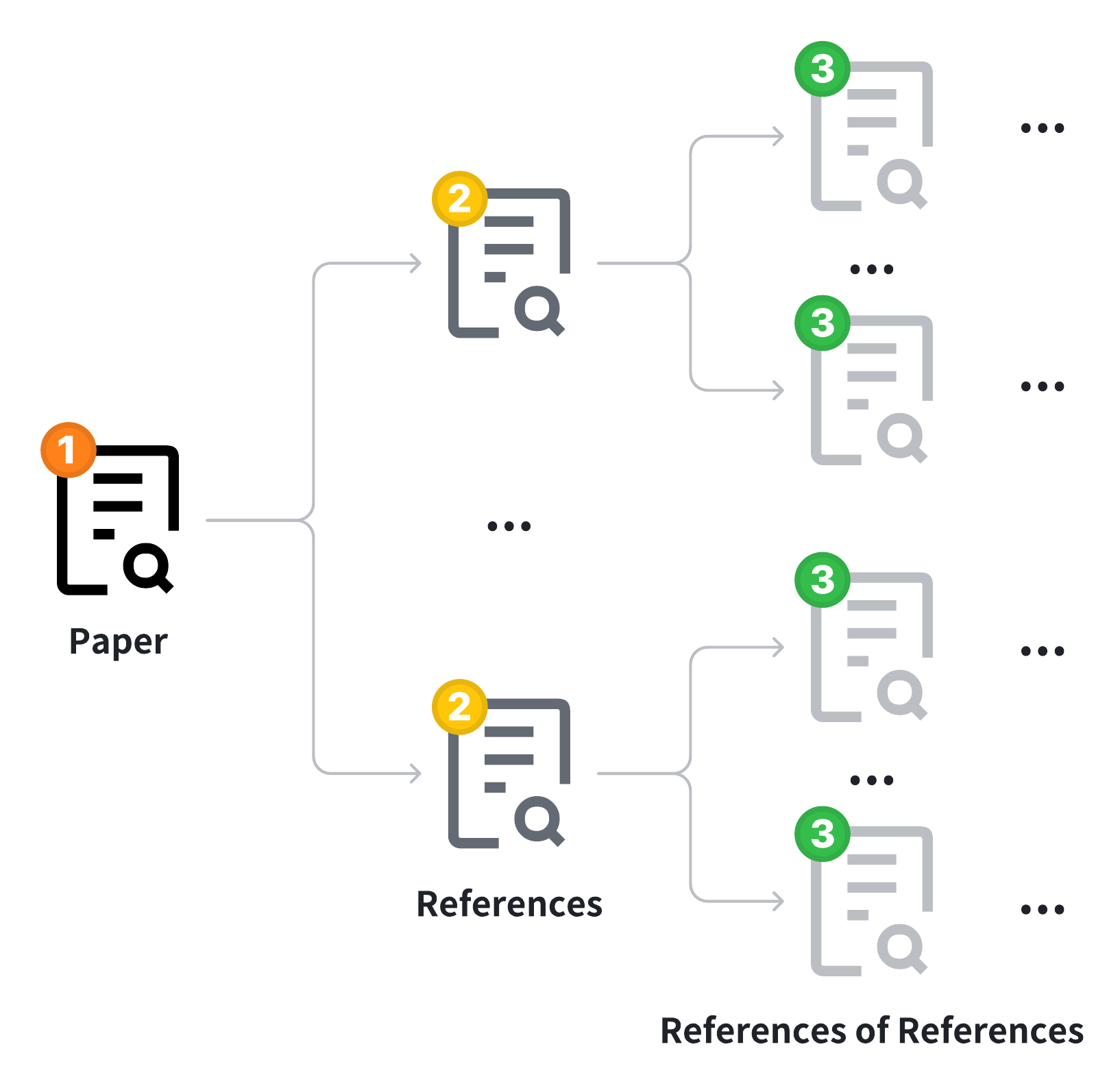}
  \caption{The architecture of RefChain.}
  \label{fig:refchain}
\end{figure}

To address this gap, we focus on modeling academic search as a recursive, citation-driven process we term the \textbf{Reference Chain} (RefChain). As illustrated in Figure~\ref{fig:refchain}, RefChain simulates how researchers follow references from one paper to another, expanding the scope of retrieval beyond direct query matches. PaSa~\cite{he2025PaSa} represents a key step in this direction, leveraging reinforcement learning (RL) to train an LLM-based agent to control RefChain expansion. However, PaSa is limited by its heavy reliance on training resources, its single-source retrieval design, and its coarse query understanding, which restrict its generalization across domains.

\begin{figure*}[htp]
  \centering
  \includegraphics[width=0.85\linewidth]{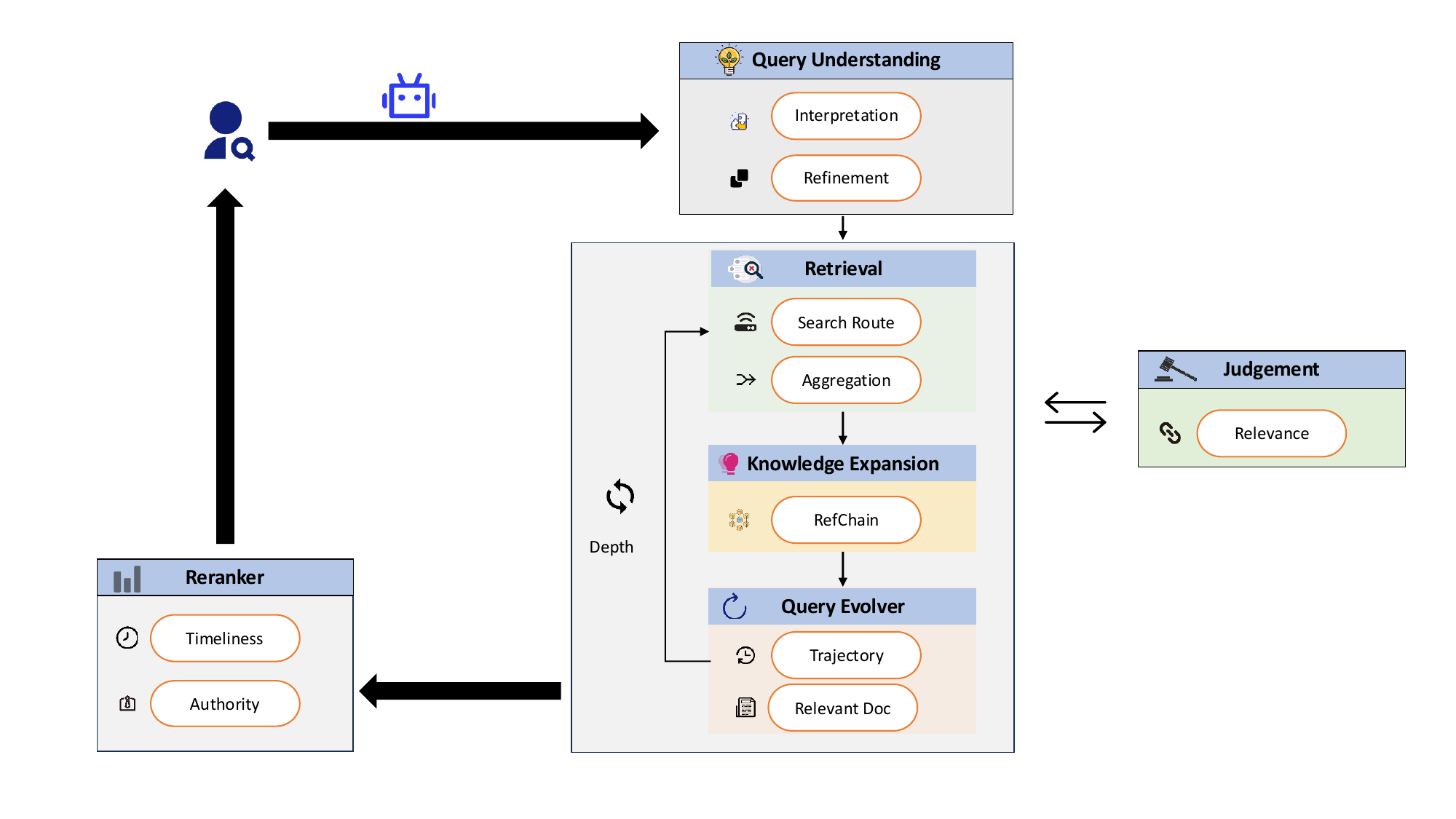}
  \caption{The overview of SPAR.}
  \label{fig:overview}
\end{figure*}

We propose \textbf{SPAR} (\textbf{S}cholar \textbf{PA}per \textbf{R}etrieval), a modular and extensible framework for academic retrieval built upon a multi-agent architecture. SPAR enhances RefChain-based exploration with five specialized components: (1) a \emph{Query Understanding Agent} that interprets domain-specific intent and refines queries accordingly; (2) a \emph{Retrieval Agent} that interfaces with multiple academic data sources; (3) a \emph{Query Evolver Agent} that performs iterative, citation-aware query reformulation; (4) a \emph{Judgement Agent} that evaluates and filters relevant papers; and (5) a \emph{Reranker Agent} that reorders retrieved results based on authority, recency, and publication quality to improve ranking effectiveness. Together, these agents support a comprehensive and dynamic academic search workflow that mirrors how human researchers conduct in-depth literature exploration (Figure~\ref{fig:overview}).

To systematically evaluate academic retrieval systems under realistic conditions, we also introduce \textbf{SPARBench}, a new benchmark comprising diverse, expert-annotated queries spanning computer science and biomedicine. Unlike existing datasets with narrow scopes, SPARBench captures the multi-faceted nature of real-world academic search. Each query and its associated relevant documents were carefully reviewed and annotated by domain experts with strong academic backgrounds, ensuring high-quality and reliable ground-truth relevance labels. This rigorous construction process makes SPARBench a robust testbed for developing and evaluating retrieval methods intended for practical academic use.

Empirical results on both AutoScholar~\cite{he2025PaSa} and SPARBench demonstrate that SPAR significantly outperforms all compared methods. On AutoScholar, SPAR achieves an F1 score of 0.3843, surpassing the previous best method, PaSa (0.2449), by 56.92\%. Notably, SPAR maintains a strong balance between recall (0.4105) and precision (0.3612), while other methods often favor one at the expense of the other. On SPARBench, SPAR is the only method that consistently achieves meaningful scores across all metrics, with an F1 of 0.3015, recall of 0.3103, and precision of 0.2932, outperforming all baselines by a clear margin. These results highlight SPAR's robustness and generalization ability across both synthetic and real-world academic search scenarios.

These findings underscore the importance of structured, agent-based retrieval frameworks for addressing the complexities of modern academic search. Our primary contributions are summarized as follows:

\begin{itemize}
    \item We propose \textbf{SPAR}, a training-free, modular, and extensible academic retrieval framework that leverages a multi-agent architecture to perform fine-grained query understanding, multi-source retrieval, RefChain-based exploration, and relevance-aware reranking.
    
    \item We introduce \textbf{SPARBench}, a high-quality, multi-domain academic retrieval benchmark featuring realistic queries and expert-annotated relevance labels across computer science and biomedicine. SPARBench enables rigorous and reproducible evaluation under practical academic search conditions.
    
    \item We conduct extensive experiments on both \textbf{AutoScholar} and \textbf{SPARBench}, demonstrating that SPAR consistently outperforms a range of strong baselines, including manual search engines (e.g., Google Scholar, Semantic Scholar), LLM-assisted retrieval pipelines, and prior agent-based methods such as PaSa and PaperFinder.
\end{itemize}

\section{Related Work}

\paragraph{Traditional Academic Search Engines}
Conventional academic search systems such as Google Scholar~\cite{vine2006google}, Semantic Scholar\cite{kinney2023semantic}, OpenAlex ~\cite{priem2022openalex}, and PubMed ~\cite{canese2013pubmed} provide effective keyword-based retrieval for well-formed queries. However, these systems rely primarily on lexical matching and are limited in their ability to handle complex, multi-intent queries~\cite{gusenbauer2020academic}. They also lack support for citation-aware exploration or semantic reasoning, which are often essential for comprehensive literature review tasks.

\paragraph{LLM-Enhanced Retrieval}
Recent advances in large language models have led to increasing interest in using LLMs to improve academic retrieval performance~\cite{zhu2023large,ma2023query}. Techniques such as query rewriting, semantic expansion, and LLM-based document re-ranking have shown promise in improving precision and recall. However, most existing approaches operate in a single-turn setting and do not support iterative, reference-driven exploration. Moreover, they rarely integrate domain-aware query understanding or multi-source retrieval strategies.

\paragraph{Agent-Based Academic Search}

Existing agent-based frameworks such as PaSa~\cite{he2025PaSa} make notable progress in automated scholarly search but remain limited by their reliance on supervised training and low modularity. To address these issues, we introduce SPAR, a training-free, modular agent framework designed for fine-grained query understanding and multi-source document exploration.

\section{Methodology}
\label{sec:method}

We introduce \textbf{SPAR} (\textbf{S}cholar \textbf{PA}per \textbf{R}etrieval), an agent-based framework for academic literature search. Given a user query, SPAR first analyzes the input to identify search intent and perform query refinement (§~\ref{sec:3.1}). It then conducts iterative retrieval via multi-source search, reference chain expansion, and query evolution (§~\ref{sec:3.2}). Finally, it re-ranks the retrieved documents based on timeliness and authority (§~\ref{sec:3.3}). An overview of the framework is shown in Figure~\ref{fig:overview}, with each component detailed in the following subsections.

\subsection{Query Interpretation and Refinement}
\label{sec:3.1}
The initial query presented by a user often represents an incomplete articulation of a complex, underlying information need, reflecting what \cite{belkin1980anomalous} termed an "Anomalous state of knowledge." Users, shaped by their unique perspectives, prior knowledge, or specific roles, naturally approach the same topic with varying informational goals and lines of inquiry \cite{teevan2005personalizing}. Therefore, effective information retrieval requires a proactive strategy to discern latent user intent and to refine the initial query into more precise and targeted instructions \cite{carpineto2012survey,croft2010search}. 
Recent studies have further emphasized the importance of query refinement in uncovering user intent, and the advent of LLMs has enabled more nuanced and context-aware query refinement techniques \cite{anand2023context,ma2023query,ye2023enhancing,liu2024query}.

SPAR incorporates a \textbf{Query Understanding} agent to interpret the user's search query and perform query refinement for subsequent precise academic paper retrieval.
Given an academic search query $q$, the agent first performs intent classification, distinguishing whether the user seeks a survey, recent advances, or methodological comparisons. It simultaneously conducts domain identification to anchor the query in a specific field of study (e.g., machine learning) and detects any temporal constraints expressed in the query (e.g., "since 2020"). These annotations help the system tailor downstream retrieval operations to the user's true research goal. Next, the agent selects one or more appropriate academic sources from a fixed set:
$\mathcal{S}=$\{Google, ArXiv, OpenAlex, Semantic Scholar, PubMed\}. 
The selection is conditioned on both the query domain and intent, ensuring source-query alignment.

The agent then determines whether the query requires multi-query refinement based on its specificity, domain clarity, and linguistic precision. If the query is broad, lacks technical terms, or includes ambiguous phrasing, the agent applies semantic disambiguation, correction, and intent-aware expansion. Refinement is guided by the query’s recognized intent and detailed refinement prompt is provided in Appendix~\ref{appendix:Prompts For Baselines}:
\begin{itemize}
    \item For survey-focused queries, the agent generates refined queries targeting different perspectives, including methods, applications, historical developments, and future challenges.
    \item For complex or specialized domains, the agent generates refined queries using domain-specific terms and technical specifications while also targeting empirical studies and primary research. 
    \item When temporal constraints are present, all refinements incorporate the specified date bounds to ensure time-sensitive relevance.
\end{itemize}

Query Understanding Agent emphasizes coverage of diverse subfields and research methodologies from the initial query. The result of this stage is a structured list of semantically enriched and disambiguated queries $\mathcal{Q}=\{q_1,q_2, \cdots, q_N\}$, each of which will be utilized to retrieve papers. This proactive query refinement lays the foundation for precise and context-aware academic paper search in SPAR.

\subsection{RefChain-based Iterative Retrieval and Query Evolution}
\label{sec:3.2}
After the Query Understanding Agent refines the query and identifies relevant sources, SPAR enters an iterative retrieval phase, coordinated by the Retrieval Agent, the Judgement Agent, and the Query Evolver Agent. The Retrieval Agent initiates this process by fetching academic papers using source-specific strategies and de-duplicating results. It also expands coverage through RefChain exploration, uncovering related work beyond the initial query matches.

The Retrieval Agent executes source-adaptive querying for each $q_i \in \mathcal{Q}$. For sources such as Semantic Scholar or OpenAlex, it extracts keywords from each query; for Google, it submits the full query string. It then consolidates results across sources by merging retrieved papers, each annotated with metadata such as title, abstract, authorship, publication date, and source.

The Judgement Agent evaluates the relevance of each retrieved paper by comparing it to the initial query and accompanying metadata. Papers scoring above the relevance threshold are added to the Related Pool $R=\{r_1,r_2,\cdots,r_m\}$. Prompts for judging relevance are provided in Appendix~\ref{appendix:prompt relevance}.

Subsequently, SPAR enhances knowledge expansion through RefChain. For each paper $r_i\in R$, the Retrieval Agent extracts its list of references either by parsing PDFs or utilizing structured metadata from sources. Then these referred papers are scored using the same Judgement Agent. High-relevance papers are merged with the Related Pool. The $K$ most relevant papers from the expanded pool are selected as final results for the current query list and stored in Paper Cache $\mathcal{P}=\{p_1,p_2,\cdots,p_K\}$. A key design decision is to limit expansion to a single RefChain layer. That is, while exploring the references of papers in the Related Pool, it does not recursively expand those references' citations. This constraint is grounded in two considerations:
\begin{itemize}
    \item \textbf{Precision and relevance: }Deeper RefChain often leads to tangential topics, reducing precision;
    \item \textbf{Computational efficiency: }Each layer significantly increases the retrieval and evaluation cost.
\end{itemize}
Compared to PaSa, which uses RL training to determine expansion depth, SPAR's deterministic, fixed-depth strategy ensures reliability, while iterative query evolution compensates by exploring new directions in a controlled manner.

To ensure depth and diversity in search results, the Query Evolver Agent then generates three new queries for $p_i\in \mathcal{P}$, focusing on its methodological insights, applications, and limitations. These queries are conditioned on the retrieval history trajectory, including the initial query, previous search queries, and the metadata of the corresponding paper. A random subset of the resulting queries is selected and added to the query list $\mathcal{Q}$ for further retrieval iterations.
Prompt for evolving query is provided in Appendix~\ref{appendix:Prompts For Query Evolution}.

This retrieval-expansion loop continues until the Paper Cache reaches a predefined size or maximum depth. To avoid redundancy, SPAR filters out previously used queries and suppresses keyword overlaps across iterations, ensuring efficient and progressive exploration of the literature space.

\subsection{Reranker}
\label{sec:3.3}
After retrieving and scoring candidate papers, a Re-ranking Module refines the final paper list. The reranking stage subsequently refines this candidate list by reordering the documents so that the most appropriate and informative items appear at the top.  
Besides the original relevance score, our reranker integrates two additional signals:  
\begin{itemize}
    \item \textbf{Publication authority}, estimated from metadata such as venue prestige and author reputation;  
    \item \textbf{Temporal relevance}, determined by whether a document satisfies explicit time constraints in the query or belongs to the most recent publications.
\end{itemize}
The prompt template that combines these factors is provided in Appendix~\ref{appendix:prompt rerank}. The final output is an ordered list of highly relevant, timely, and authoritative academic papers tailored to the user’s intent.

\section{SPARBench}

\begin{figure*}[htp]
  \centering
  \includegraphics[width=0.8\linewidth]{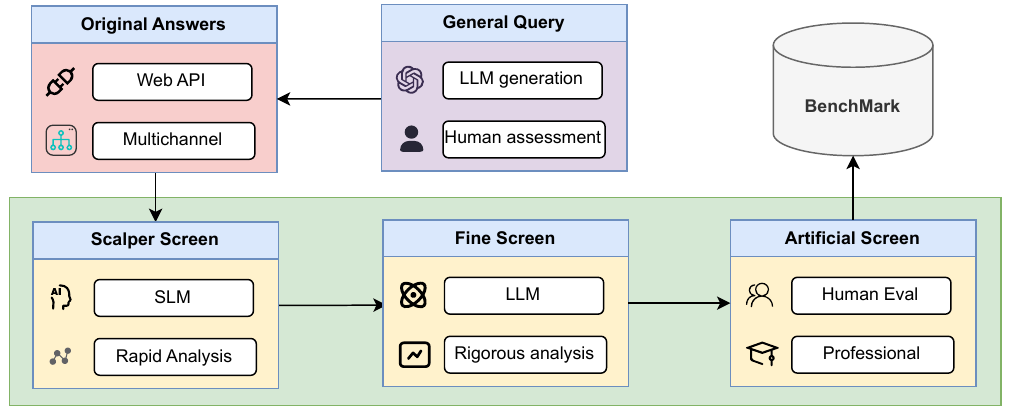}
  \caption{SPARBench construction pipeline. The process includes expert-curated seed queries, GPT-4o-based query expansion, multi-source document retrieval, and a three-stage relevance filtering procedure combining language models and expert annotation.}

  \label{fig:benchcmark-pipeline}
\end{figure*}

Despite growing interest in scholarly information retrieval, the field still lacks robust and standardized benchmarks for systematic and realistic evaluation. This absence limits reproducibility and hinders progress in developing generalizable academic search systems.

Existing resources remain limited in both scope and quality. For example, AutoScholar~\cite{he2025PaSa} is a synthetic dataset constructed from AI conference papers between 2023 and 2024. Although it pairs GPT-4o-generated queries with relevant documents, only 100 query-document pairs were manually reviewed, raising concerns about label quality and applicability to real-world scenarios. Another benchmark, RealScholarQuery~\cite{he2025PaSa}, contains 50 expert-written queries collected post-hoc from AI researchers, introducing potential evaluation bias toward models tuned for that specific setup.

Most prior benchmarks focus on closed corpora~\cite{ajith2024litsearch,voorhees2021trec,cohan2004specter}, using static queries and documents. Such settings fail to capture key aspects of academic search, including query understanding, multi-source retrieval, and reference-based exploration. Despite efforts toward more comprehensive evaluation, no existing benchmark supports end-to-end assessment encompassing ranking, reasoning, source selection, and iterative exploration.

To address these limitations, we introduce \textbf{SPARBench}, a benchmark for evaluating academic retrieval systems under realistic conditions. Unlike previous efforts, SPARBench draws from multiple academic sources—including arXiv, PubMed, OpenAlex, and Semantic Scholar—covering diverse disciplines such as computer science and biomedicine.

SPARBench reflects natural academic search behavior. Initial queries are generated by GPT-4o and then rigorously filtered by domain experts. The dataset includes multi-intent queries with incomplete grammar and minor spelling errors to simulate real-world user input. Relevance judgments follow a multi-stage process combining automatic filtering, small and large language models, and manual validation by experts, ensuring high label quality and domain fidelity.

Given the high cost of producing high-quality academic retrieval data, the current version includes 50 carefully curated queries, each undergoing expert review. This initial release prioritizes depth and reliability, providing a solid foundation for future extensions to broader domains and larger query sets. SPARBench will be publicly released to support further research in academic search.

\subsection{Benchmark Characteristics}

\begin{itemize}
    \item \textbf{Realistic Queries:} Simulate authentic academic search behavior through multi-intent, semantically rich queries with incomplete grammar and minor spelling errors.
    
    \item \textbf{Cross-Domain Coverage:} Supports evaluation across computer science and biomedicine, enabling assessment of domain-general and domain-specific retrieval capabilities.
    
    \item \textbf{Multi-Source Corpus:} Integrates documents from arXiv, PubMed, OpenAlex, and Semantic Scholar to reduce source-specific bias and improve retrieval realism.
    
    \item \textbf{High-Quality Annotations:} A multi-stage labeling pipeline combines LLM-based filtering with expert validation, ensuring high-quality annotations and domain consistency.

\end{itemize}

\subsection{Construction Method}

Figure~\ref{fig:benchcmark-pipeline} outlines the multi-stage pipeline used to construct SPARBench. A set of seed queries was manually curated based on real academic research scenarios. These were expanded using GPT-4o~\cite{hurst2024gpt} to introduce linguistic and semantic diversity. After expert screening, 50 queries were selected—35 from computer science and 15 from biomedicine.

Each query was submitted independently to arXiv, PubMed, OpenAlex, and Semantic Scholar, producing an initial candidate set of 198K documents. Relevance assessment proceeded in three stages:

\begin{enumerate}
    \item \textbf{Initial Pruning:} Coarse relevance was estimated using Qwen2.5-7B-Instruct~\cite{qwen2.5}, reducing the set to 3K candidates.
    \item \textbf{Refinement:} Qwen2.5-72B-Instruct~\cite{qwen2.5} performed fine-grained filtering, yielding 2K documents.
    \item \textbf{Expert Validation:} Graduate-level computer science annotators manually reviewed the remaining candidates, selecting approximately 560 relevant documents (averaging 12 per query).
\end{enumerate}

The final benchmark comprises 50 queries and 560 expert-verified relevant documents. Stage-wise statistics are reported in Appendix~\ref{appendix:benchmark-stage-volume} (Figure~\ref{appendix:fig:benchmark-stage-volume}).

SPARBench fills a critical gap in academic retrieval research by offering a realistic, high-quality benchmark tailored for end-to-end evaluation of scholarly search systems.

\section{Experiments}

\subsection{Evaluation Setup}

We evaluated our method against a diverse set of baselines, including traditional academic and web search engines, as well as LLM-enhanced retrieval systems. The evaluated baselines include:

\begin{itemize}
    \item \textsc{Google (G):} Standard Google search using the original query.
    \item \textsc{Google+GPT-4o (G+GPT):}Query rewritten for clarity using GPT-4o~\cite{hurst2024gpt} before Google search.
    \item \textsc{Google Scholar (GS):}Direct retrieval from Google Scholar without LLM intervention.
    \item \textsc{ChatGPT Search (CS):}We Submit query to ChatGPT, which is powered by search-enabled GPT-4o.
    \item \textsc{Google-ArXiv (GA):}Google search restricted to arXiv.org.
    \item \textsc{Google-ArXiv + LLM (GA+LLM):} Query refined using  LLM  before Google search restricted by arXiv.
    \item \textsc{OpenAlex+LLM (OA+LLM):}Keywords extracted by LLM for the retrieval of the OpenAlex API.
    \item \textsc{Semantic Scholar+LLM(2S+LLM):}LLM-extracted keywords used for the Semantic Scholar search.
    \item \textsc{PubMed+LLM(PM+LLM):}LLM-generated keywords for PubMed searches.
    \item \textsc{PaSa:}An LLM-driven academic search agent optimized through reinforcement learning~\cite{he2025PaSa}.
    \item \textsc{PaperFinder:}An LLM-powered academic search assistant accessed at <https://paperfinder.allen.ai/chat>(accessed July 9, 2025). It mimics human-like iterative literature search by decomposing queries, tracking citations, and providing relevance explanations.~\cite{ai2_paper_finder_2025}

\end{itemize}

For all "+LLM" variants, we use Qwen3-32B~\cite{qwen3} for keyword extraction, relevance estimation or query refinement. For "+GPT" variants, GPT-4o is employed to rewrite the query for improved clarity before web search. Task-specific prompting strategies are detailed in Appendix~\ref{appendix:Prompts For Baselines}.

We evaluate retrieval performance using three standard metrics: \textbf{Precision}, \textbf{Recall}, and \textbf{F1}, computed at the document level for each query. Importantly, each retrieval system operates over its own native search source (e.g., OpenAlex, Semantic Scholar, Google Scholar), rather than performing search on a shared benchmark corpus. This setup reflects realistic usage scenarios and allows for end-to-end evaluation of each system's full retrieval pipeline, including query understanding, source selection, search execution, and result ranking. Our goal is to assess the overall effectiveness of each system as a holistic academic search solution.

Let $TP$ be the number of true positives (relevant documents correctly retrieved), $FP$ the number of false positives (irrelevant documents retrieved), and $FN$ the number of false negatives (relevant documents not retrieved). The metrics are defined as follows:

\begin{align}
    \text{Precision} &= \frac{TP}{TP + FP} \\
    \text{Recall} &= \frac{TP}{TP + FN} \\
    \text{F1} &= \frac{2 \cdot \text{Precision} \cdot \text{Recall}}{\text{Precision} + \text{Recall}}
\end{align}

\textbf{Precision} measures the proportion of retrieved documents that are truly relevant, reflecting retrieval accuracy. \textbf{Recall} measures the proportion of relevant documents that are successfully retrieved, reflecting coverage. \textbf{F1} is the harmonic mean of Precision and Recall, providing a balanced assessment of both accuracy and completeness.

We report results on two benchmarks:

\begin{itemize}
    \item \textbf{AutoScholar:} A synthetic benchmark introduced in the PaSa paper, designed to evaluate retrieval precision on fine-grained AI domain queries.
    \item \textbf{SPARBench:} Our curated benchmark featuring real-world queries from computer science and biomedicine, with expert-validated relevance annotations.
\end{itemize}

\subsection{Main Result}

\begin{table*}[htp]
    \centering
    \small
    \begin{tabular}{c|ccc|ccc}
    \toprule
    \multirow{2}{*}{\textbf{Method}} & \multicolumn{3}{c|}{\textbf{AutoScholar}} & \multicolumn{3}{c}{\textbf{SPARBench}} \\
    & \textbf{F1} & \textbf{Recall } & \textbf{Precision} & \textbf{F1} & \textbf{Recall} & \textbf{Precision}  \\
    \midrule
    G & - & 0.2015 & -  & - & 0.000 & -  \\
    G+GPT & - & 0.2683 & - & 0.0092 & 0.0082 & 0.0106 \\
    GS & - & 0.1130 & - & 0.0043 & 0.0038 & 0.0050  \\
    CS & 0.0869 &  0.3046 & 0.0507  & 0.0045 & 0.0038 & 0.0055 \\
    GA & 0.0400 & 0.1571 & 0.0229 & 0.2451 & 0.2800 & 0.2180 \\
    GA+LLM & 0.0556 & 0.1692 & 0.0333 & 0.1923 & 0.1613 & 0.2382  \\
    PM+LLM & - & 0.000 & - & - & 0.000 & -  \\
    OA+LLM & 0.0045 & 0.1083 & 0.0023 & 0.0242 & 0.0988 & 0.0138  \\
    2S+LLM & 0.0044 & 0.0833 & 0.0023  & 0.0135 & 0.0449 & 0.0080  \\
    PaSa & 0.2449 & 0.7931 & 0.1448  & 0.1041 & 0.1009 & 0.1076  \\
    PaperFinder & 0.0506 & \textbf{0.8333} & 0.0261  & 0.0418 & 0.1474 & 0.0244  \\
    \midrule
    \textbf{SPAR (ours)} & \textbf{0.3843} & 0.4105 & \textbf{0.3612} & \textbf{0.3015} & \textbf{0.3103} & \textbf{0.2932} \\
    \bottomrule
    \end{tabular}
    \caption{
        Comparison of retrieval performance across different methods on the AutoScholar and SPARBench benchmarks. “–” indicates metrics unavailable due to missing valid document.
        }
    \label{tab:main-results}
\end{table*}

As shown in Table~\ref{tab:main-results}, our proposed method \textsc{SPAR} consistently outperforms all baselines across both benchmarks. On the \textsc{AutoScholar} dataset, SPAR achieves the highest F1 score of \textbf{0.3843} and the highest precision of \textbf{0.3612}, while maintaining a competitive recall (\textbf{0.4105}). This demonstrates its strong ability to retrieve relevant documents with high accuracy and balance. On \textsc{SPARBench}, SPAR also surpasses all other methods, achieving the best F1 score of \textbf{0.3015}, recall of \textbf{0.3103}, and precision of \textbf{0.2932}. In contrast, prior methods such as \textsc{GA+LLM}, \textsc{PaSa}, and \textsc{PaperFinder} exhibit either lower precision or significant performance imbalance (e.g., high recall but very low precision). Notably, while \textsc{PaperFinder} obtains the highest recall (0.8333) on AutoScholar, its precision (0.0261) is extremely low, leading to a much lower F1 score. These results highlight SPAR’s superior capability in balancing precision and recall, thereby providing robust and effective academic document retrieval across diverse settings.

\section{Analysis and Discussion}

\subsection{Effects on Query Interpretation}
\label{sec:Effects on Query Interpretation}

\begin{table}[ht]
    \centering
    \small
    \begin{tabular*}{\columnwidth}{@{\extracolsep{\fill}}lcccc@{}}
    \toprule
    \textbf{Benchmark} & \textbf{QInterp} & \textbf{F1} & \textbf{Recall} & \textbf{Precision} \\
    \midrule
    \multirow{2}{*}{AutoScholar}
        & w    & \textbf{0.19} & 0.24 & \textbf{0.16} \\
        & w/o & 0.18 & \textbf{0.25} & 0.14 \\
    \midrule
    \multirow{2}{*}{SPARBench}
        & w   & \textbf{0.22} & 0.16 & \textbf{0.34} \\
        & w/o & 0.21 & \textbf{0.21} & 0.21 \\
    \bottomrule
    \end{tabular*}
    \caption{
        Effect of query interpretation (QInterp) on retrieval performance across benchmarks.
    }
    \label{table:query_interpretation}
\end{table}

Query Interpretation (QInterp) enhances retrieval by analyzing the query intent, selecting appropriate sources, and performing intent-aware rewriting. As described in Appendix~\ref{appendix:QUERY interp pipeline}, this module introduces structural awareness that enables better alignment between queries and target documents.

Table~\ref{table:query_interpretation} reports retrieval results with and without QInterp across two benchmarks. On both datasets, enabling QInterp improves overall F1 and precision. In SPARBench, precision increases substantially from 0.21 to 0.34, reflecting improved ranking relevance in a complex, multi-source environment. However, recall tends to decrease (e.g., 0.25 to 0.24 in AutoScholar, 0.21 to 0.16 in SPARBench), likely due to more restrictive interpretations that favor precision over coverage. This trade-off is consistent with observations in baseline systems employing aggressive query rewriting (Table~\ref{tab:main-results}).

These results suggest that query interpretation is beneficial for precision-oriented retrieval, especially in settings requiring fine-grained query understanding and source selection. Future work may explore hybrid strategies that balance interpretation with recall-aware expansion.

\subsection{Impact of RefChain}
\label{sec:Impact of RefChain}

The RefChain mechanism significantly improves document recall by expanding the set of candidates through citation-based traversal. As shown in Table~\ref{tab:refchain-impact} (Appendix~\ref{appendix:table-refchain-impact}), RefChain enhances recall-oriented metrics on both the AutoScholar and SPARBench benchmarks.  

In AutoScholar, RefChain increases the recall after similarity filtering from 0.41 to 0.44 and raw recall from 0.58 to 0.77, while the average retrieved documents rise from 306.9 to 569.1. Similarly, on SPARBench, recall improves from 0.13 to 0.15, the raw recall from 0.26 to 0.31, and the retrieval volume from 367.8 to 504.9.  
However, this recall improvement reduces precision due to increased noise. In AutoScholar, precision drops from 0.29 to 0.19, and in SPARBench from 0.22 to 0.16. 

These results indicate that RefChain is most beneficial in recall-critical scenarios, such as retrieval-augmented generation (RAG) for academic synthesis. In contrast, precision-focused retrieval systems may prefer to disable RefChain to minimize noise and reduce downstream filtering overhead.

\subsection{Benefits of Query Evolution}
\label{section:analysis:Benefits of Query Evolution}

Query Evolution refines search queries by leveraging retrieval history and high-relevance documents, enhancing search focus via semantic guidance from top-ranked results (see case study in Appendix~\ref{appendix:case study:Query Evolution vs Native method}). 

Table~\ref{tab:queryevolver_performance} reports its impact on F1, recall, and precision across two academic search benchmarks: \textbf{AutoScholar} and \textbf{SPARBench}.

\begin{table}[ht]
    \small
    \begin{tabular*}{\columnwidth}{@{\extracolsep{\fill}}lcccc@{}}
    \toprule
    \textbf{Benchmark} & \textbf{Evolution} & \textbf{F1} & \textbf{Recall} & \textbf{Precision} \\
    \midrule
    \multirow{2}{*}{AutoScholar} 
    & w & \textbf{0.34} & 0.41 & \textbf{0.29} \\
    & w/o & 0.33 & \textbf{0.43} & 0.27 \\
    \midrule
    \multirow{2}{*}{SPARBench} 
    & w  & \textbf{0.26} & \textbf{0.24} & \textbf{0.27} \\
    & w/o  & 0.24 & \textbf{0.24} & 0.25 \\
    \bottomrule
    \end{tabular*}
    \caption{Impact of query evolution on retrieval performance. \textbf{Evolution} denotes application of Query Evolution.}
    \label{tab:queryevolver_performance}
\end{table}

Query Evolution consistently improves F1 in both benchmarks. Precision increases by 0.02 in each dataset, indicating more targeted retrieval. Although recall slightly decreases in AutoScholar, the overall shift toward higher precision demonstrates the effectiveness of focused querying. Prompts used for query evolution are provided in Appendix~\ref{appendix:Prompts For Query Evolution}.

\subsection{Reranking Strategy and Its Advantages}

The reranking module reorders the top-10 retrieved documents to optimize \textbf{Recall@5}. Table~\ref{tab:rerank_effects} reports its effects on two benchmarks. On AutoScholar, Recall@5 increases from 0.3146 to 0.4015, representing a 27.6\% relative improvement. On SPARBench, Recall@5 improves from 0.1588 to 0.1662, a 4.7\% relative gain. 

These results demonstrate the module's effectiveness in prioritizing authoritative and contextually relevant documents, thereby enhancing retrieval quality and user experience. The larger improvement observed on AutoScholar suggests that reranking is particularly beneficial for datasets characterized by simpler query structures.

\begin{table}[ht]
    \small
    \centering
    \begin{tabular*}{\columnwidth}{@{\extracolsep{\fill}}lcc@{}}
    \toprule
    \textbf{Benchmark} & \textbf{Reranking} & \textbf{Recall@5} \\
    \midrule
    \multirow{2}{*}{AutoScholar} 
    & w/o & 0.3146 \\
    & w   & \textbf{0.4015} \\
    \midrule
    \multirow{2}{*}{SPARBench} 
    & w/o & 0.1588 \\
    & w   & \textbf{0.1662} \\
    \bottomrule
    \end{tabular*}
    \caption{Effect of reranking on Recall@5 for the top 5 retrieved documents across benchmarks.}
    \label{tab:rerank_effects}
\end{table}

\subsection{Relevance Assessment: Model and Prompt Selection}
\label{sec:analysis-rel-assessment}

We examine how model and prompt choices affect relevance assessment performance. Larger models do not consistently outperform smaller ones. A systematic evaluation reveals the most effective configuration.

We compare two prompt styles—\textit{brief} and \textit{complex}—across seven language models on the AutoScholar and SPARBench benchmarks (Appendix~\ref{appendix:prompt relevance}; Table~\ref{tab:performance_with_different_relevance_assesment}). On AutoScholar, Qwen3-32B with the brief prompt achieves the highest F1 score (0.38). On SPARBench, LLaMA3.3-70B~\cite{grattafiori2024llama} with the same prompt performs best (F1: 0.30). Based on overall performance across both datasets, we select Qwen3-32B (brief) as the default configuration.

Additional generalization experiments (Appendix~\ref{appendix:table-performance_with_different_relevance_assesment_onepsource}) confirm the robustness of this choice, underscoring the importance of prompt design and model selection in relevance assessment.

\section{Conclusion}

We present \textbf{SPAR}, a modular multi-agent framework for academic paper retrieval, designed to tackle the challenges of underspecified queries, fragmented sources, and evolving information needs. SPAR consists of four key stages: (1) a \textbf{Query Understanding} agent that interprets user intent through intent classification, domain detection, and temporal constraint parsing, followed by intent-aware query refinement; (2) an \textbf{Iterative Retrieval} phase that integrates source-adaptive querying and \textbf{RefChain}-based citation expansion for recall-oriented exploration; (3) a \textbf{Query Evolver} agent that diversifies search trajectories by generating follow-up queries based on previously retrieved papers; and (4) a \textbf{Reranker} that ranks results using relevance, timeliness, and publication authority.

To evaluate SPAR, we construct \textbf{SPARBench}, a benchmark of semantically complex academic queries with expert-labeled relevance. Experiments on SPARBench and AutoScholar demonstrate that SPAR consistently outperforms strong baselines. It achieves an F1 score of \textbf{0.3843} on AutoScholar, a \textbf{+56\% improvement over PaSa}, and \textbf{0.3015} on SPARBench, the only method delivering balanced performance across all metrics. 

Our results validate the synergy of symbolic planning (RefChain), LLM-powered query evolution, and agent-based modular design in addressing the complexity of scholarly retrieval tasks. SPAR and SPARBench offer a reproducible and extensible foundation for advancing intelligent academic search systems.

\section*{Limitations}

Despite the strong performance of SPAR, several limitations remain. 

First, SPAR limits citation-based expansion (RefChain) to a single traversal depth. While this design reduces latency and suppresses noise, it may miss deeply nested but highly relevant works, especially in long citation chains that characterize foundational research.

Second, RefChain substantially improves recall but introduces noisy candidates, leading to lower precision. This trade-off, while acceptable for recall-oriented tasks, may be suboptimal in scenarios that demand high-precision retrieval, such as targeted literature reviews.

Third, SPAR currently relies on static prompting and rule-based orchestration. The lack of feedback-driven learning or user interaction modeling hinders personalization and adaptation over time. Incorporating reinforcement signals or retrieval-based supervision could make the system more robust in dynamic search environments.

Finally, although SPARBench provides a valuable testbed for semantically complex academic queries, it remains limited in scale and domain diversity. Future work should extend SPARBench to cover additional disciplines and query types, enabling broader generalization and facilitating standardized evaluation for next-generation academic search systems.

\bibliography{acl}

\begin{thebibliography}{27}
\providecommand{\natexlab}[1]{#1}

\bibitem[{Achiam et~al.(2023)Achiam, Adler, Agarwal, Ahmad, Akkaya, Aleman, Almeida, Altenschmidt, Altman, Anadkat et~al.}]{achiam2023gpt}
Josh Achiam, Steven Adler, Sandhini Agarwal, Lama Ahmad, Ilge Akkaya, Florencia~Leoni Aleman, Diogo Almeida, Janko Altenschmidt, Sam Altman, Shyamal Anadkat, and 1 others. 2023.
\newblock Gpt-4 technical report.
\newblock \emph{arXiv preprint arXiv:2303.08774}.

\bibitem[{Ajith et~al.(2024)Ajith, Xia, Chevalier, Goyal, Chen, and Gao}]{ajith2024litsearch}
Anirudh Ajith, Mengzhou Xia, Alexis Chevalier, Tanya Goyal, Danqi Chen, and Tianyu Gao. 2024.
\newblock Litsearch: A retrieval benchmark for scientific literature search.
\newblock \emph{arXiv preprint arXiv:2407.18940}.

\bibitem[{{Allen Institute for AI}(2025)}]{ai2_paper_finder_2025}
{Allen Institute for AI}. 2025.
\newblock Ai2 paper finder: {LLM}-powered academic search assistant.
\newblock \url{https://paperfinder.allen.ai/chat}.
\newblock Accessed: 2025-07-09.

\bibitem[{Anand et~al.(2023)Anand, Setty, Anand et~al.}]{anand2023context}
Abhijit Anand, Vinay Setty, Avishek Anand, and 1 others. 2023.
\newblock Context aware query rewriting for text rankers using llm.
\newblock \emph{arXiv preprint arXiv:2308.16753}.

\bibitem[{Belkin(1980)}]{belkin1980anomalous}
Nicholas~J Belkin. 1980.
\newblock Anomalous states of knowledge as a basis for information retrieval.
\newblock \emph{Canadian journal of information science}, 5(1):133--143.

\bibitem[{Canese and Weis(2013)}]{canese2013pubmed}
Kathi Canese and Sarah Weis. 2013.
\newblock Pubmed: the bibliographic database.
\newblock \emph{The NCBI handbook}, 2(1):2013.

\bibitem[{Carpineto and Romano(2012)}]{carpineto2012survey}
Claudio Carpineto and Giovanni Romano. 2012.
\newblock A survey of automatic query expansion in information retrieval.
\newblock \emph{Acm Computing Surveys (CSUR)}, 44(1):1--50.

\bibitem[{Cohan et~al.(2004)Cohan, Feldman, Beltagy, Downey, and Weld}]{cohan2004specter}
A~Cohan, S~Feldman, I~Beltagy, D~Downey, and DS~Weld. 2004.
\newblock Specter: document-level representation learning using citation-informed transformers. 2020.
\newblock \emph{arXiv preprint arXiv:2004.07180}.

\bibitem[{Croft et~al.(2010)Croft, Metzler, and Strohman}]{croft2010search}
W~Bruce Croft, Donald Metzler, and Trevor Strohman. 2010.
\newblock \emph{Search engines: Information retrieval in practice}, volume 520.
\newblock Addison-Wesley Reading.

\bibitem[{Grattafiori et~al.(2024)Grattafiori, Dubey, Jauhri, Pandey, Kadian, Al-Dahle, Letman, Mathur, Schelten, Vaughan et~al.}]{grattafiori2024llama}
Aaron Grattafiori, Abhimanyu Dubey, Abhinav Jauhri, Abhinav Pandey, Abhishek Kadian, Ahmad Al-Dahle, Aiesha Letman, Akhil Mathur, Alan Schelten, Alex Vaughan, and 1 others. 2024.
\newblock The llama 3 herd of models.
\newblock \emph{arXiv preprint arXiv:2407.21783}.

\bibitem[{Gusenbauer and Haddaway(2020)}]{gusenbauer2020academic}
Michael Gusenbauer and Neal~R Haddaway. 2020.
\newblock Which academic search systems are suitable for systematic reviews or meta-analyses? evaluating retrieval qualities of google scholar, pubmed, and 26 other resources.
\newblock \emph{Research synthesis methods}, 11(2):181--217.

\bibitem[{He et~al.(2025)He, Huang, Feng, Lin, Zhang, Li et~al.}]{he2025PaSa}
Yichen He, Guanhua Huang, Peiyuan Feng, Yuan Lin, Yuchen Zhang, Hang Li, and 1 others. 2025.
\newblock Pasa: An llm agent for comprehensive academic paper search.
\newblock \emph{arXiv preprint arXiv:2501.10120}.

\bibitem[{Hurst et~al.(2024)Hurst, Lerer, Goucher, Perelman, Ramesh, Clark, Ostrow, Welihinda, Hayes, Radford et~al.}]{hurst2024gpt}
Aaron Hurst, Adam Lerer, Adam~P Goucher, Adam Perelman, Aditya Ramesh, Aidan Clark, AJ~Ostrow, Akila Welihinda, Alan Hayes, Alec Radford, and 1 others. 2024.
\newblock Gpt-4o system card.
\newblock \emph{arXiv preprint arXiv:2410.21276}.

\bibitem[{Kinney et~al.(2023)Kinney, Anastasiades, Authur, Beltagy, Bragg, Buraczynski, Cachola, Candra, Chandrasekhar, Cohan et~al.}]{kinney2023semantic}
Rodney Kinney, Chloe Anastasiades, Russell Authur, Iz~Beltagy, Jonathan Bragg, Alexandra Buraczynski, Isabel Cachola, Stefan Candra, Yoganand Chandrasekhar, Arman Cohan, and 1 others. 2023.
\newblock The semantic scholar open data platform.
\newblock \emph{arXiv preprint arXiv:2301.10140}.

\bibitem[{Liu et~al.(2024)Liu, Feng, Xue, Wang, Wu, Lu, Zhao, Deng, Zhang, Ruan et~al.}]{liu2024deepseek}
Aixin Liu, Bei Feng, Bing Xue, Bingxuan Wang, Bochao Wu, Chengda Lu, Chenggang Zhao, Chengqi Deng, Chenyu Zhang, Chong Ruan, and 1 others. 2024.
\newblock Deepseek-v3 technical report.
\newblock \emph{arXiv preprint arXiv:2412.19437}.

\bibitem[{Liu and Mozafari(2024)}]{liu2024query}
Jie Liu and Barzan Mozafari. 2024.
\newblock Query rewriting via large language models.
\newblock \emph{arXiv preprint arXiv:2403.09060}.

\bibitem[{Ma et~al.(2023)Ma, Gong, He, Zhao, and Duan}]{ma2023query}
Xinbei Ma, Yeyun Gong, Pengcheng He, Hai Zhao, and Nan Duan. 2023.
\newblock Query rewriting in retrieval-augmented large language models.
\newblock In \emph{Proceedings of the 2023 Conference on Empirical Methods in Natural Language Processing}, pages 5303--5315.

\bibitem[{Priem et~al.(2022)Priem, Piwowar, and Orr}]{priem2022openalex}
Jason Priem, Heather Piwowar, and Richard Orr. 2022.
\newblock Openalex: A fully-open index of scholarly works, authors, venues, institutions, and concepts.
\newblock \emph{arXiv preprint arXiv:2205.01833}.

\bibitem[{Team et~al.(2023)Team, Anil, Borgeaud, Alayrac, Yu, Soricut, Schalkwyk, Dai, Hauth, Millican et~al.}]{team2023gemini}
Gemini Team, Rohan Anil, Sebastian Borgeaud, Jean-Baptiste Alayrac, Jiahui Yu, Radu Soricut, Johan Schalkwyk, Andrew~M Dai, Anja Hauth, Katie Millican, and 1 others. 2023.
\newblock Gemini: a family of highly capable multimodal models.
\newblock \emph{arXiv preprint arXiv:2312.11805}.

\bibitem[{Teevan et~al.(2005)Teevan, Dumais, and Horvitz}]{teevan2005personalizing}
Jaime Teevan, Susan~T Dumais, and Eric Horvitz. 2005.
\newblock Personalizing search via automated analysis of interests and activities.
\newblock In \emph{Proceedings of the 28th annual international ACM SIGIR conference on Research and development in information retrieval}, pages 449--456.

\bibitem[{Vine(2006)}]{vine2006google}
Rita Vine. 2006.
\newblock Google scholar.
\newblock \emph{Journal of the Medical Library Association}, 94(1):97.

\bibitem[{Voorhees et~al.(2021)Voorhees, Alam, Bedrick, Demner-Fushman, Hersh, Lo, Roberts, Soboroff, and Wang}]{voorhees2021trec}
Ellen Voorhees, Tasmeer Alam, Steven Bedrick, Dina Demner-Fushman, William~R Hersh, Kyle Lo, Kirk Roberts, Ian Soboroff, and Lucy~Lu Wang. 2021.
\newblock Trec-covid: constructing a pandemic information retrieval test collection.
\newblock In \emph{ACM SIGIR Forum}, volume~54, pages 1--12. ACM New York, NY, USA.

\bibitem[{Yang et~al.(2025{\natexlab{a}})Yang, Li, Yang, Zhang, Hui, Zheng, Yu, Gao, Huang, Lv, Zheng, Liu, Zhou, Huang, Hu, Ge, Wei, Lin, Tang, Yang, Tu, Zhang, Yang, Yang, Zhou, Zhou, Lin, Dang, Bao, Yang, Yu, Deng, Li, Xue, Li, Zhang, Wang, Zhu, Men, Gao, Liu, Luo, Li, Tang, Yin, Ren, Wang, Zhang, Ren, Fan, Su, Zhang, Zhang, Wan, Liu, Wang, Cui, Zhang, Zhou, and Qiu}]{qwen3}
An~Yang, Anfeng Li, Baosong Yang, Beichen Zhang, Binyuan Hui, Bo~Zheng, Bowen Yu, Chang Gao, Chengen Huang, Chenxu Lv, Chujie Zheng, Dayiheng Liu, Fan Zhou, Fei Huang, Feng Hu, Hao Ge, Haoran Wei, Huan Lin, Jialong Tang, and 41 others. 2025{\natexlab{a}}.
\newblock Qwen3 technical report.
\newblock \emph{arXiv preprint arXiv:2505.09388}.

\bibitem[{Yang et~al.(2025{\natexlab{b}})Yang, Li, Yang, Zhang, Hui, Zheng, Yu, Gao, Huang, Lv et~al.}]{yang2025qwen3}
An~Yang, Anfeng Li, Baosong Yang, Beichen Zhang, Binyuan Hui, Bo~Zheng, Bowen Yu, Chang Gao, Chengen Huang, Chenxu Lv, and 1 others. 2025{\natexlab{b}}.
\newblock Qwen3 technical report.
\newblock \emph{arXiv preprint arXiv:2505.09388}.

\bibitem[{Yang et~al.(2024)Yang, Yang, Zhang, Hui, Zheng, Yu, Li, Liu, Huang, Wei, Lin, Yang, Tu, Zhang, Yang, Yang, Zhou, Lin, Dang, Lu, Bao, Yang, Yu, Li, Xue, Zhang, Zhu, Men, Lin, Li, Xia, Ren, Ren, Fan, Su, Zhang, Wan, Liu, Cui, Zhang, and Qiu}]{qwen2.5}
An~Yang, Baosong Yang, Beichen Zhang, Binyuan Hui, Bo~Zheng, Bowen Yu, Chengyuan Li, Dayiheng Liu, Fei Huang, Haoran Wei, Huan Lin, Jian Yang, Jianhong Tu, Jianwei Zhang, Jianxin Yang, Jiaxi Yang, Jingren Zhou, Junyang Lin, Kai Dang, and 22 others. 2024.
\newblock Qwen2.5 technical report.
\newblock \emph{arXiv preprint arXiv:2412.15115}.

\bibitem[{Ye et~al.(2023)Ye, Fang, Li, and Yilmaz}]{ye2023enhancing}
Fanghua Ye, Meng Fang, Shenghui Li, and Emine Yilmaz. 2023.
\newblock Enhancing conversational search: Large language model-aided informative query rewriting.
\newblock \emph{arXiv preprint arXiv:2310.09716}.

\bibitem[{Zhu et~al.(2023)Zhu, Yuan, Wang, Liu, Liu, Deng, Chen, Liu, Dou, and Wen}]{zhu2023large}
Yutao Zhu, Huaying Yuan, Shuting Wang, Jiongnan Liu, Wenhan Liu, Chenlong Deng, Haonan Chen, Zheng Liu, Zhicheng Dou, and Ji-Rong Wen. 2023.
\newblock Large language models for information retrieval: A survey.
\newblock \emph{arXiv preprint arXiv:2308.07107}.

\end{thebibliography}

\newpage
\appendix

\section{Prompt Template}

\label{appendix:prompt collections}

\subsection{Prompt For Baseline}
\label{appendix:Prompts For Baselines}

\begin{tcolorbox}[
      colback=white!10!white,
      colframe=black!75!black,
      title=\textbf{Prompt for Query Refinement},
      fonttitle=\bfseries,
      breakable,
      sharp corners=south
    ]
    \label{appendix:prompt-baseline_prompt_query_refine}

    Generate a search query suitable for Google based on the given academic paper-related query. Please adhere to the following instructions:
    
    \begin{enumerate}
      \item \textbf{Understand the Query}: Carefully read and comprehend the given academic query.
      \item \textbf{Identify Key Elements}: Extract the main research domain, specific methods, or core concepts.
      \item \textbf{Formulate the Search Query}: Construct a concise and effective query that captures these components and is suitable for academic search engines.
      \item \textbf{Avoid Site Constraints}: Do \textbf{not} include any site-specific filters (e.g., \texttt{site:xxx}).
      \item \textbf{Output Format}: Only generate the refined query using the format below.
    \end{enumerate}
    
    \textbf{[User’s Query]}: \textcolor{blue}{\textbf{\{UserQuery\}}}\\
    \textbf{[Generated Search Query]}: <your query here>
\end{tcolorbox}

\begin{tcolorbox}[
      colback=white!10!white,
      colframe=black!75!black,
      title=\textbf{Prompt for keywords extraction},
      fonttitle=\bfseries,
      breakable,
      sharp corners=south
    ]
    \label{appexdix:baseline_prompt_keywords_extraction}
    Extract optimal search keywords from the given research question, specifically optimized for the \textcolor{blue}{\textbf{\texttt{\{source\}}}} academic database. Your task is to generate concise, comma-separated query terms that will maximize relevant paper retrieval on this platform.
    
    \textbf{Source-Specific Guidelines:}
    \begin{itemize}
        \item \textbf{Semantic Scholar:}
        \begin{itemize}
            \item Focus on technical terminology and core concepts.
            \item Include methodological terms.
            \item Consider author-centric keywords if prominent researchers are known.
            \item Emphasize computer science and AI terminology where relevant.
        \end{itemize}
        
        \item \textbf{OpenAlex:}
        \begin{itemize}
            \item Prioritize broader academic terms.
            \item Include interdisciplinary connections.
            \item Balance specificity with coverage.
            \item Include field classifications where relevant.
        \end{itemize}
    
        \item \textbf{PubMed:}
        \begin{itemize}
            \item Emphasize medical/biological terminology.
            \item Include relevant MeSH (Medical Subject Headings) terms.
            \item Consider clinical and biomedical contexts.
            \item Include chemical/drug names or biological processes where relevant.
        \end{itemize}
    \end{itemize}

    \textbf{Response Format:}\\
    \texttt{[Start]} keyword1, keyword2, keyword3, ... \texttt{[End]}
    
    \textbf{Examples by Source:}
    \begin{itemize}
        \item Semantic Scholar: [Start] transformer architecture, attention mechanism, language model fine-tuning [End]
        \item OpenAlex: [Start] neural networks, deep learning, artificial intelligence, pattern recognition [End]
        \item PubMed: [Start] CRISPR-Cas9, gene editing, genetic therapy, chromosomal modification [End]
    \end{itemize}
    
    Now, extract optimized search keywords for \textcolor{blue}{\textbf{\texttt{\{source\}}}} from this question:\textcolor{blue}{\textbf{\texttt{\{user\_query\}}}}
    
\end{tcolorbox}

\subsection{Prompt For Query Evolution}
\label{appendix:Prompts For Query Evolution}

\begin{tcolorbox}[
    colback=white!10!white,
    colframe=black!75!black,
    breakable]
    \label{prompt for query evolution}
    You are an academic search expert helping explore a research topic more thoroughly.\\
    
    \#\#\# CONTEXT:\\
    - Original Query: \textcolor{blue}{\textbf{\{user\_query}\}}\\
    - Previously Searched Queries: \textcolor{blue}{\textbf{\{searched\_queries\}}}\\
    - Relevant Document Title: \textcolor{blue}{\textbf{\{doc\_title}\}}\\
    - Document Abstract: \textcolor{blue}{\textbf{\{doc\_abstract}\}}\\
    - Document Field: \textcolor{blue}{\textbf{\{doc\_field}\}}\\
    
    \#\#\# TASK:\\
    Generate \textcolor{blue}{\textbf{\{N\}}} NEW search queries that explore different aspects of this research area:\\
    
    1. A query exploring METHODOLOGICAL alternatives or comparisons\\
    2. A query focusing on APPLICATIONS or implementations\\
    3. A query addressing LIMITATIONS, challenges, or critiques\\
    
    Each query should be:\\
    - Clearly different from previously searched queries\\
    - Based on insights from the document\\
    - Relevant to the original research question\\
    - Specific enough to retrieve focused results\\
    
    \#\#\# IMPORTANT NOTE:\\
    If document information is missing or insufficient (e.g., empty abstract), generate queries based primarily on the original query and your knowledge of the research domain. Focus on exploring complementary aspects of the topic rather than requiring specific document details.\\
    
    \#\#\# OUTPUT FORMAT:\\
    Return a JSON array of strings containing only the expanded queries:\\
    \textbf{[Query 1,Query 2, Query 3]}

\end{tcolorbox}

\subsection{Prompt for Relevance}
\label{appendix:prompt relevance}

\begin{tcolorbox}[
    colback=white!10!white,
    colframe=black!75!black,
    title=prompt for Relevance -- Brief,
    breakable]
    You are an expert in academic research. Given a query and a document in the context of a scholarly paper search, evaluate their relevance on a scale from 0 to 1, where 0 means completely irrelevant and 1 means highly relevant. Base your evaluation on the query’s intent, key concepts, and the document’s content. Provide a score and explain your reasoning consistently.
    
    Query: \textcolor{blue}{\textbf{\{UserQuery\}}}\\
    Document:\\
    Title: \textcolor{blue}{\textbf{\{title\}}}\\
    Abstract: \textcolor{blue}{\textbf{\{abstract\}}}\\
    
    \textbf{Score}: [Your score between 0 and 1]\\
    \textbf{Reasoning}: [Your explanation]
\end{tcolorbox}

\begin{tcolorbox}[
    colback=white!10!white,
    colframe=black!75!black,
    title=prompt for Relevance -- Complex,
    breakable]
    \textbf{You are a rigorous and highly discerning academic search relevance evaluator.} Your task is to critically assess the relationship between the user's query and the provided scholarly article. Apply a strict, high-standard academic lens to evaluate conceptual alignment, topical focus, and methodological relevance. Be skeptical of superficial keyword matches or loosely related themes. Only assign a high relevance score (on a 0–1 scale) when there is clear and substantial alignment in research purpose, methods, and contribution. \textbf{Err on the side of conservatism in scoring—precision and selectivity are paramount.}
    
    \vspace{0.5em}
    \textbf{Input Format} \\
    Query: Raw academic search query \\
    Article:
    \begin{itemize}
        \item Title: Academic article title
        \item Abstract: Abstract text summarizing the paper's content
    \end{itemize}
    
    \vspace{0.5em}
    \textbf{Hierarchical Evaluation Protocol}
    \begin{enumerate}
        \item \textbf{Critical Relevance Check (Binary Gate)} \\
        If the document contains \emph{zero} of the following, automatically score 0.0:
        \begin{itemize}
            \item Core subject keywords from query
            \item Matching research domain
            \item Thematic alignment with query intent
        \end{itemize}
        
        \item \textbf{Detailed Scoring Criteria (Only if passes Critical Check)}
        
        \textbf{A. Core Topic Alignment (0–0.6)}
        \begin{itemize}
            \item 0.5–0.6: Directly addresses primary subject with matching terminology
            \item 0.3–0.4: Related subfield but different focus area
            \item 0.1–0.2: Only tangential connection through peripheral terms
            \item 0.0: Fails Critical Relevance Check
        \end{itemize}
        
        \textbf{B. Contextual Precision (0–0.3)}
        \begin{itemize}
            \item 0.2–0.3: Explicitly addresses query's specific technical aspects
            \item 0.1: General thematic similarity without concrete details
            \item 0.0: No meaningful connection to query intent
        \end{itemize}
        
        \textbf{C. Depth Validation (0–0.1)}
        \begin{itemize}
            \item 0.1: Provides experimental validation/novel theoretical framework
            \item 0.05: Mentions concept without substantive analysis
            \item 0.0: Superficial treatment of subject
        \end{itemize}
    \end{enumerate}
    
    \vspace{0.5em}
    \textbf{Scoring Matrix (Sum Components A + B + C)}
    \begin{itemize}
        \item 0.00–0.19: Completely irrelevant / off-topic
        \item 0.20–0.39: Minimal relevance — shares domain but different focus
        \item 0.40–0.59: Partial relevance — addresses some aspects
        \item 0.60–0.79: Substantial relevance — covers key elements
        \item 0.80–1.00: Optimal match — comprehensive coverage
    \end{itemize}
    
    \vspace{0.5em}
    \textbf{Anti-Gaming Rules}
    \begin{itemize}
        \item Penalize -0.3 for keyword stuffing without contextual relevance
        \item Penalize -0.2 for misleading titles/abstracts
        \item If score $<$ 0.4, round down to nearest 0.1
        \item If score $\geq 0.7$, require positive marks in all 3 criteria
    \end{itemize}
    
    \vspace{0.5em}
    \textbf{Examples}
    
    \textbf{Example 1 (Low Score)} \\
    \textit{Query:} ``Machine learning for early Alzheimer's diagnosis using MRI'' \\
    \textit{Article:} ``Statistical analysis of MRI machine calibration errors'' \\
    \textit{Reasoning:} Fails Critical Relevance — no ML or Alzheimer's content \\
    \textit{Score:} 0.15
    
    \vspace{0.5em}
    \textbf{Example 2 (High Score)} \\
    \textit{Query:} ``Federated learning optimization in IoT networks'' \\
    \textit{Article:} ``Adaptive Gradient Compression for Energy-Efficient Federated Learning in Edge Computing Environments'' \\
    \textit{Reasoning:} Directly addresses FL optimization (0.6) + technical specifics (0.25) + experimental validation (0.1) \\
    \textit{Score:} 0.86
    
    \vspace{0.5em}
    \textbf{Input Data} \\
    \textbf{Query:} \textcolor{blue}{\textbf{\{query\}}} \\
    \textbf{Article:} \textcolor{blue}{\textbf{\{doc\}}}
    
    \vspace{0.5em}
    \textbf{Output Format} \\
    \textbf{Reasoning:} [Concise technical justification] \\
    \textbf{Score:} [0.00–1.00]
\end{tcolorbox}

\subsection{Prompt For Reranking}
\label{appendix:prompt rerank}

\begin{tcolorbox}[
    colback=white!10!white,
    colframe=black!75!black,
    title=Reranking with Time Requirement,
    breakable]
    
    Please rerank the following \textcolor{blue}{\textbf{\{N\}}} academic papers in response to the query:\textcolor{blue}{\textbf{\{Query\}}}\\
    
    \textbf{Consider these factors in your reranking:}
    \begin{enumerate}
        \item \textbf{Authority:}
        \begin{itemize}
            \item Publication venue prestige (top conferences/journals rank higher)
            \item Author prominence (authors with higher h-index or citation counts rank higher)
        \end{itemize}
        \item \textbf{Timeliness:}
        \begin{itemize}
            \item \textcolor{red}{\textit{The query specifically asks for recent/current papers, so strongly prefer newer papers}}
        \end{itemize}
        \item \textbf{Maintain reasonable relevance to the original query}
    \end{enumerate}
    
    \textbf{For each paper, provide:}
    \begin{enumerate}
        \item A new numerical rank (1 being the highest)
        \item A brief justification (1-2 sentences)
        \item A new relevance score between 0-1 that incorporates both relevance and the factors above
    \end{enumerate}
    
    \textbf{List of papers with original relevance scores (title, year, venue, authors, relevance):}\\\\
    \textcolor{blue}{\textbf{\{Doc List Here\}}}\\
    
    \textbf{Please provide your reranking with new scores and concise justifications in the following format for each document:}\\
    Document [index]: [score] - [justification]\\
    
    \textbf{For example}:\\
    Document 1: 9.5 - Highly relevant as it directly addresses the query topic with empirical evidence.\\
    Document 2: 7.0 - Somewhat relevant but focuses on a tangential aspect of the query.\\
\end{tcolorbox}

\begin{tcolorbox}[
    colback=white!10!white,
    colframe=black!75!black,
    title=Reranking without Time Requirement,
    breakable]
    Please rerank the following \textcolor{blue}{\textbf{\{N\}}} academic papers in response to the query: \textcolor{blue}{\textbf{\{Query\}}}\\
    
    \textbf{Consider these factors in your reranking:}
    \begin{enumerate}
        \item \textbf{Authority:}
        \begin{itemize}
            \item Publication venue prestige (top conferences/journals rank higher)
            \item Author prominence (authors with higher h-index or citation counts rank higher)
        \end{itemize}
        \item \textbf{Timeliness:}
        \begin{itemize}
            \item\textcolor{red}{\textit{Generally prefer more recent papers, but don't overly penalize influential older papers}}
        \end{itemize}
        \item \textbf{Maintain reasonable relevance to the original query}
    \end{enumerate}
    
    \textbf{For each paper, provide:}
    \begin{enumerate}
        \item A new numerical rank (1 being the highest)
        \item A brief justification (1-2 sentences)
        \item A new relevance score between 0-1 that incorporates both relevance and the factors above
    \end{enumerate}
    
    \textbf{List of papers with original relevance scores (title, year, venue, authors, relevance):}\\\\
    \textcolor{blue}{\textbf{\{Doc List Here\}}}\\
    
    \textbf{Please provide your reranking with new scores and concise justifications in the following format for each document:}\\
    Document [index]: [score] - [justification]\\
    
    \textbf{For example}:\\
    Document 1: 9.5 - Highly relevant as it directly addresses the query topic with empirical evidence.\\
    Document 2: 7.0 - Somewhat relevant but focuses on a tangential aspect of the query.\\
    
\end{tcolorbox}

\section{BenchMark Information}

\subsection{SPARBench Example}
\label{appendix:sparbench_example}

\begin{tcolorbox}[
    colback=white!10!white,
    colframe=black!75!black,
    title=An Example of SPARBench,
    breakable]
    
    \textbf{Question:} 
    "What are the potentials and ethical challenges of gene editing technologies (e.g., CRISPR) in treating genetic diseases? Provide specific explanations and recent research progress."
    
    \vspace{0.5em}
    \textbf{Source Metadata:}
    \begin{itemize}[leftmargin=*,noitemsep]
        \item \textbf{Search Time:} 2025-04-10
    \end{itemize}
    
    \begin{itemize}[leftmargin=*,noitemsep]
    \item \textbf{Reference Answers:}
    \begin{enumerate}[leftmargin=*,noitemsep]
        \item 
        \begin{itemize}[leftmargin=*]
            \item \textbf{Paper ID:} \url{http://genome.cshlp.org/content/24/9/1526.full.pdf}
            \item \textbf{Title:} "Seamless gene correction of $\beta$-thalassemia mutations in patient-specific iPSCs using CRISPR/Cas9 and \textit{piggyBac}"
            \item \textbf{Abstract:} $\beta$-thalassemia, one of the most common genetic diseases worldwide, is caused by mutations in human hemoglobin beta (HBB) gene. Creation of induced pluripotent stem cells (iPSCs) from $\beta$-thalassemia patients could offer an approach to cure this disease. Correction of disease-causing mutations in iPSCs can restore normal function and provide a rich source for transplantation. In this study, we used the latest gene-editing tool, CRISPR/Cas9 technology, combined with piggyBac transposon to efficiently correct patient-derived mutations without leaving any residual footprint. No off-target effects were detected in corrected iPSCs, which retain full pluripotency and normal karyotypes. When differentiated into erythroblasts using monolayer culture, gene-corrected cells restored HBB expression compared to the parental line. Our study provides an effective footprint-free correction method, thereby demonstrating a critical step toward future application of cell-based gene therapy for monogenic diseases.
            \item \textbf{Authors:} Fei Xie, Lin Ye, Judy C. Chang, Ashley I. Beyer, Jiaming Wang, Marcus O. Muench, Yuet Wai Kan
            \item \textbf{Year:} 2014
            \item \textbf{Citation Count:} 381
            \item \textbf{Source:} OpenAlex
            \item \textbf{Similarity Scores:} 
                \begin{itemize}
                    \item Small Model: 0.7
                    \item LLM: 0.85
                \end{itemize}
        \end{itemize}
        \item 
        \begin{itemize}[leftmargin=*]
            \item \textbf{Paper ID:} \url{https://www.nature.com/articles/s41392-019-0089-y.pdf}
            \item \textbf{Title:}  "Applications of genome editing technology in the targeted therapy of human diseases: mechanisms, advances and prospects",
            \item \textbf{Abstract:} "..."
            
        \end{itemize}
        
        \item [...] 
    \end{enumerate}
    \end{itemize}
\end{tcolorbox}

\section{Strategy Result}

\subsection{Query Interpretation Result}
\label{appendix:query_interpretation_result}

\begin{tcolorbox}[
    colback=white!10!white,
    colframe=black!75!black,
    title=Example of Query Interpretation,
    breakable]
    \begin{description}
        \item[User Query:] What improvements are needed in vaccine development efficiency to respond to emerging infectious diseases? Provide a multi-angle analysis.
        \item[Query Intent:] Literature review and multi-angle analysis
        \item[Domain:] Biomedical sciences / Vaccine development
        \item[Suitable Sources:] pubmed, openalex
        \item[Needs Expansion:] true
        \item[Expansion Reason:] The query is somewhat broad and could be improved by incorporating specific technical terms or focusing on particular aspects of vaccine development. Terms like 'vaccine development efficiency' are relatively general and may yield a wide range of results, some of which might not be directly relevant to responding to emerging infectious diseases. The query could benefit from specifying methodologies (e.g., computational modeling, platform technologies, mRNA-based approaches), clarifying the scope (e.g., preclinical vs clinical stages), and defining the type of research sought (e.g., reviews, meta-analyses, case studies). Including terms such as 'platform vaccines', 'reverse vaccinology', or 'rapid antigen identification' would enhance precision and relevance within the domain of biomedical sciences.
        
        \item[Expanded Queries:]
        
        \begin{itemize}
            \item Systematic review of vaccine development methods for emerging infectious diseases
            \item Literature review of vaccine application strategies in global health emergencies (2021-2025)
            \item Historical survey of vaccine innovation in response to pandemics since 2000
            \item Comparative analysis of future challenges in accelerating vaccine development post-pandemic
            \item Survey of multi-disciplinary approaches to improve vaccine design efficiency
        \end{itemize}
        
        \item[Time Requirement Description:] NO
        \item[Source Reason:] PubMed is the most suitable source for this query due to its focus on biomedical and life sciences research, which directly aligns with vaccine development. OpenAlex can also be useful as it provides interdisciplinary context and broader metadata, supporting a multi-angle analysis without time constraints.
    \end{description}
    \label{query_interpretation_result}
\end{tcolorbox}

\subsection{Query Evolution vs Native Method}
\label{appendix:case study:Query Evolution vs Native method}
This example illustrates query-document semantic relationships via color coding. Tokens highlighted in the same color indicate shared or closely related concepts between queries and document content.
\begin{tcolorbox}[
    colback=white!10!white,
    colframe=black!75!black,
    title=Query Evolution vs Native method,
    breakable]
    \textbf{Original Query:} "Can you identify any papers that analysed the use of target networks with \textcolor{green}{\textbf{linear function approximation}}, needed in theoretical properties of target networks?"
    
    \textbf{Previously Searched Queries:} 
    \begin{itemize}[leftmargin=*,noitemsep,topsep=0pt]
        \item Survey of target networks with \textcolor{green}{\textbf{linear function approximation}} methods
        \item Systematic review of historical development of target networks with \textcolor{green}{\textbf{linear function approximation}}
        \item Literature review of target networks applications using \textcolor{green}{\textbf{linear function approximation}}
        \item Can you identify any papers that analysed the use of target networks with \textcolor{green}{\textbf{linear function approximation}}, needed in theoretical properties of target networks?
        \item State-of-the-art in theoretical properties of target networks using \textcolor{green}{\textbf{linear function approximation}} (2024-2025)
        \item Comparative analysis of future challenges in target networks with \textcolor{green}{\textbf{linear function approximation}}
    \end{itemize}
        
    \textbf{Relevant Document Title:} "A Unifying View of \textcolor{green}{\textbf{Linear Function Approximation}} in Off-Policy \textcolor{green}{\textbf{RL}} Through Matrix Splitting and \textcolor{red}{\textbf{Preconditioning}}"
        
    \textbf{Document Abstract:} 
    "Traditionally, \textcolor{blue}{\textbf{TD}} and \textcolor{blue}{\textbf{FQI}} are viewed as differing in the number of updates toward the target value function: \textcolor{blue}{\textbf{TD}} makes one update, \textcolor{blue}{\textbf{FQI}} makes an infinite number, and \textcolor{blue}{\textbf{Partial Fitted Q-Iteration (PFQI)}} performs a finite number, such as the use of a target network in Deep Q-Networks (DQN) in the \textcolor{red}{\textbf{off-policy}} setting. This perspective, however, fails to capture the \textcolor{red}{\textbf{convergence}} connections between these algorithms and may lead to incorrect conclusions, for example, that the \textcolor{red}{\textbf{convergence}} of \textcolor{blue}{\textbf{TD}} implies the \textcolor{red}{\textbf{convergence}} of \textcolor{blue}{\textbf{FQI}}. In this paper, we focus on \textcolor{green}{\textbf{linear value function approximation}} and offer a new perspective, unifying \textcolor{blue}{\textbf{TD}}, \textcolor{blue}{\textbf{FQI}}, and \textcolor{blue}{\textbf{PFQI}} as the same iterative method for solving the Least Squares Temporal Difference (LSTD) system, but using different \textcolor{red}{\textbf{preconditioners}} and matrix splitting schemes. \textcolor{blue}{\textbf{TD}} uses a constant \textcolor{red}{\textbf{preconditioner}}, \textcolor{blue}{\textbf{FQI}} employs a data-feature adaptive \textcolor{red}{\textbf{preconditioner}}, and \textcolor{blue}{\textbf{PFQI}} transitions between the two. Then, we reveal that in the context of \textcolor{green}{\textbf{linear function approximation}}, increasing the number of updates under the same target value function essentially represents a transition from using a constant \textcolor{red}{\textbf{preconditioner}} to data-feature adaptive \textcolor{red}{\textbf{preconditioner}}. This unifying perspective also simplifies the analyses of the \textcolor{red}{\textbf{convergence conditions}} for these algorithms and clarifies many issues. Consequently, we fully characterize the \textcolor{red}{\textbf{convergence}} of each algorithm without assuming specific properties of the chosen features (e.g., linear independence). We also examine how common assumptions about feature representations affect \textcolor{red}{\textbf{convergence}}, and discover new conditions on features that are important for \textcolor{red}{\textbf{convergence}}. These \textcolor{red}{\textbf{convergence conditions}} allow us to establish the \textcolor{red}{\textbf{convergence connections}} between these algorithms and to address important questions."
    
    \textbf{- Document Field:} "cs.LG"\\
        
    \textbf{QueryEvolution:} 
    \begin{itemize}[leftmargin=*,noitemsep,topsep=0pt]
        \item Practical implementation of \textcolor{blue}{\textbf{Partial Fitted Q-Iteration}} with \textcolor{green}{\textbf{linear function approximation}} in \textcolor{red}{\textbf{off-policy}} \textcolor{green}{\textbf{reinforcement learning}} settings
        \item Challenges and \textcolor{red}{\textbf{convergence}} limitations of using fixed versus adaptive \textcolor{red}{\textbf{preconditioners}} in target network-based \textcolor{green}{\textbf{reinforcement learning}} algorithms
    \end{itemize}
        
    \textbf{NativeMethod:}
        \begin{itemize}[leftmargin=*,noitemsep,topsep=0pt]
            \item Real-world implementations and case studies of target networks using \textcolor{green}{\textbf{linear function approximation}}
            \item Critique of \textcolor{red}{\textbf{convergence}} and stability issues in target networks employing \textcolor{green}{\textbf{linear function approximation}}
        \end{itemize}
    
\end{tcolorbox}

\subsection{RefChain Effect Details}
As shown in Table~\ref{tab:refchain-impact}, using RefChain can improve the final Recall by 7.32\% and 15.38\% on AutoScholar and SPARBench.
\label{appendix:table-refchain-impact}

\begin{table*}[htp]
    \centering
    \small
    \begin{tabular}{lcccccc}
    \toprule
    Benchmark & RefChain & Recall & Precision & Raw Doc Num & Valid Doc Num & Recall (Raw) \\
    \midrule
    \multirow{2}{*}{AutoScholar}
        & w/o & 0.41 & \textbf{0.29} & 306.90 & 3.95 & 0.58 \\
        & w    & \textbf{0.44} & 0.19 & \textbf{569.08} & \textbf{6.58} & \textbf{0.77} \\
    \midrule
    \multirow{2}{*}{SPARBench}
        & w/o  & 0.13 & \textbf{0.22} & 367.81 & 10.77 & 0.26 \\
        & w    & \textbf{0.15} & 0.16 & \textbf{504.94} & \textbf{15.00} & \textbf{0.31} \\
    \bottomrule
    \end{tabular}
    \caption{
        Impact of RefChain on document retrieval metrics across two benchmarks.
        Enabling RefChain improves recall but introduces more noise, leading to a drop in precision.
        \textbf{Recall} and \textbf{Precision} are computed based on documents retained after relevance filtering.
        \textbf{Raw Doc Num} refers to the total number of documents retrieved before filtering;
        \textbf{Valid Doc Num} indicates the number of relevant documents identified after filtering;
        \textbf{Recall (Raw)} is recall calculated over the full set of raw retrieved documents.
    }
    \label{tab:refchain-impact}
\end{table*}

\subsection{Performance of Varies Relevance Assesment}
\subsubsection{Performance on Benchmarks}
Table~\ref{tab:performance_with_different_relevance_assesment} shows the comparison of different models for relevance judgment. Qwen3-32B with brief instruction achieves the best F1 score of 0.38.
\label{appendix:table-performance_with_different_relevance_assesment}

\begin{table*}[htp]
    \centering
    \small
    \begin{tabular}{lccc|ccc}
    \hline
     & \multicolumn{3}{c}{\textbf{AutoScholar}} & \multicolumn{3}{c}{\textbf{SPARBench}} \\
    \textbf{Model and Inst Variant} & F1 & Recall & Precision & F1 & Recall & Precision \\
    \hline
    PaSa\_Selector & 0.34 & 0.41 & 0.29 & 0.26 & 0.24 & 0.27 \\
    \hline
    Llama3.1-8B (brief) & 0.11 & 0.19 & 0.08 & 0.20 & 0.18 & 0.23 \\
    Llama3.1-8B (complex) & 0.13 & 0.48 & 0.07 & 0.23 & 0.24 & 0.22 \\
    Llama3.3-70B (brief) & 0.20 & 0.38 & 0.13 & \textbf{0.30}& 0.31 & \textbf{0.29} \\
    Llama3.3-70B (complex) & 0.18 & 0.34 & 0.12 & 0.17 & 0.34 & 0.12 \\
    \hline
    Qwen2.5-7B (brief) & 0.34 & 0.47 & 0.27 & 0.28 & 0.28 & 0.28 \\
    Qwen2.5-7B (complex) & 0.08 & 0.09 & 0.06 & 0.24 & 0.27 & 0.21 \\
    Qwen2.5-72B (brief) & 0.33 & 0.51 & 0.24 & 0.24 & \textbf{0.38} & 0.17 \\
    Qwen2.5-72B (complex) & 0.17 & 0.44 & 0.11 & 0.19 & 0.27 & 0.15 \\
    \hline
    Qwen3-8B (brief) & 0.29 & \textbf{0.53} & 0.20 & 0.25 & 0.30 & 0.22 \\
    Qwen3-8B (complex) & 0.31 & 0.45 & 0.24 & 0.23 & 0.31 & 0.19 \\
    Qwen3-14B (brief) & 0.21 & 0.37 & 0.14 & 0.25 & 0.32 & 0.20 \\
    Qwen3-14B (complex) & 0.22 & 0.38 & 0.15 & 0.23 & 0.34 & 0.17 \\
    \rowcolor{gray!20}
    Qwen3-32B (brief) & \textbf{0.38} & 0.41 & \textbf{0.36} & 0.24 & 0.29 & 0.21 \\
    Qwen3-32B (complex) & 0.08 & 0.29 & 0.05 & 0.18 & 0.30 & 0.12 \\
    \hline
    \end{tabular}
    \caption{Performance Comparison of Different Models on AutoScholar and SPARBench Datasets(where 'brief' refers to inst-brief and 'complex' to inst-complex). Prompt details can be found in Appendix~\ref{appendix:prompt relevance}. }
    \label{tab:performance_with_different_relevance_assesment}
\end{table*}

\subsubsection{Performance on OpenSource Dataset}
\label{appendix:table-performance_with_different_relevance_assesment_onepsource}
Table~\ref{tab:f1-scores-grouped-opensource} shows the comparison of different models for relevance judgment on three open-source benchmarks. Qwen3-32B with brief instruction achieves the best average performance among all benchmarks.
\begin{table*}[htp]
    \centering
    \small
    \begin{tabular}{lccc}
    \toprule
    \textbf{Model and Inst Variant} & \textbf{TREC-Covid} & \textbf{Scidocs} & \textbf{LitSearch} \\
    \midrule
    PaSa\_Selector & 0.7010 & 0.1291 & 0.4980 \\
    \midrule
    LLaMA3.1-8B (brief) & 0.6967 & 0.7453 & 0.5695 \\
    LLaMA3.1-8B (complex) & 0.6537 & 0.5251 & 0.5080 \\
    LLaMA3.3-70B (brief) & 0.7047 & 0.7366 & \textbf{0.5737} \\
    LLaMA3.3-70B (complex) & 0.6942 & 0.3278 & 0.5108 \\
    \midrule
    Qwen2.5-7B (brief) & 0.6930 & 0.3022 & 0.4808 \\
    Qwen2.5-7B (complex) & 0.6693 & 0.0751 & 0.3571 \\
    Qwen2.5-72B (brief) & 0.7163 & \textbf{0.7715} & 0.5830 \\
    Qwen2.5-72B (complex) & 0.6921 & 0.1668 & 0.4374 \\
    \midrule
    Qwen3-8B (brief) & 0.7143 & 0.3553 & 0.5224 \\
    Qwen3-8B (complex) & 0.6569 & 0.1203 & 0.4335 \\
    Qwen3-14B (brief) & 0.7170 & 0.4756 & 0.5338 \\
    Qwen3-14B (complex) & 0.6853 & 0.1238 & 0.4481 \\
    \rowcolor{gray!20}
    Qwen3-32B (brief) & \textbf{0.7256 }& 0.6082 & 0.5566 \\
    Qwen3-32B (complex) & 0.6729 & 0.1651 & 0.4550 \\
    \bottomrule
    \end{tabular}
    \caption{F1 scores of various models on TREC-Covid~\cite{voorhees2021trec}, Scidocs~\cite{cohan2004specter}, and LitSearch-NLP-Class~\cite{ajith2024litsearch} datasets. }
    \label{tab:f1-scores-grouped-opensource}
\end{table*}

\section{Query Interpretation Overview}
\label{appendix:QUERY interp pipeline}
\begin{figure*}[htp]
  \centering
  \includegraphics[width=\linewidth]{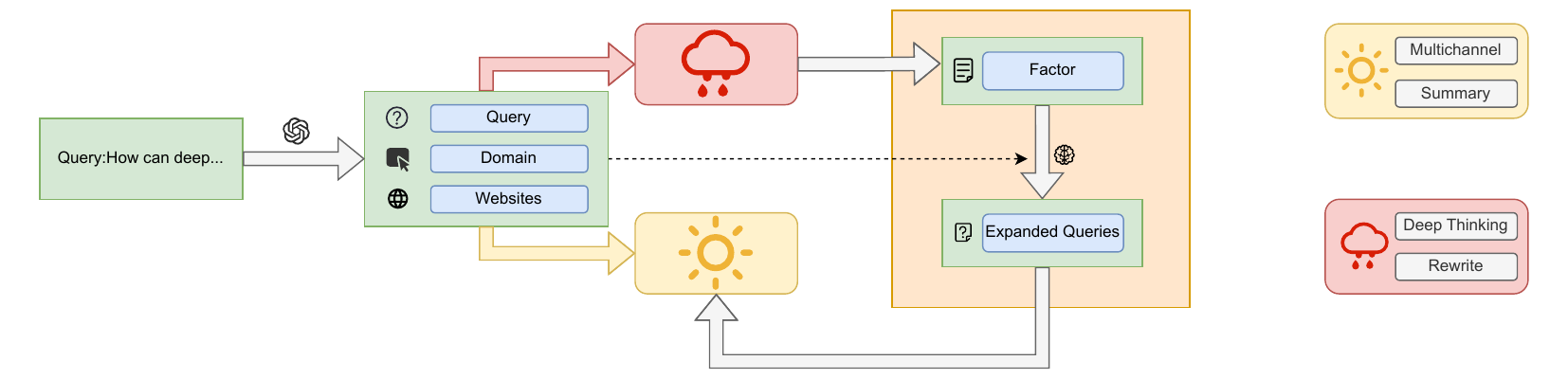}
  \caption{The Overview of Query Interpretation Module}
  \label{fig:QUERY interp pipeline}
\end{figure*}

\section{SPARBench Stage Volume Change}
\label{appendix:benchmark-stage-volume}

\begin{figure*}[htp]
  \centering
  \includegraphics[width=0.7\linewidth]{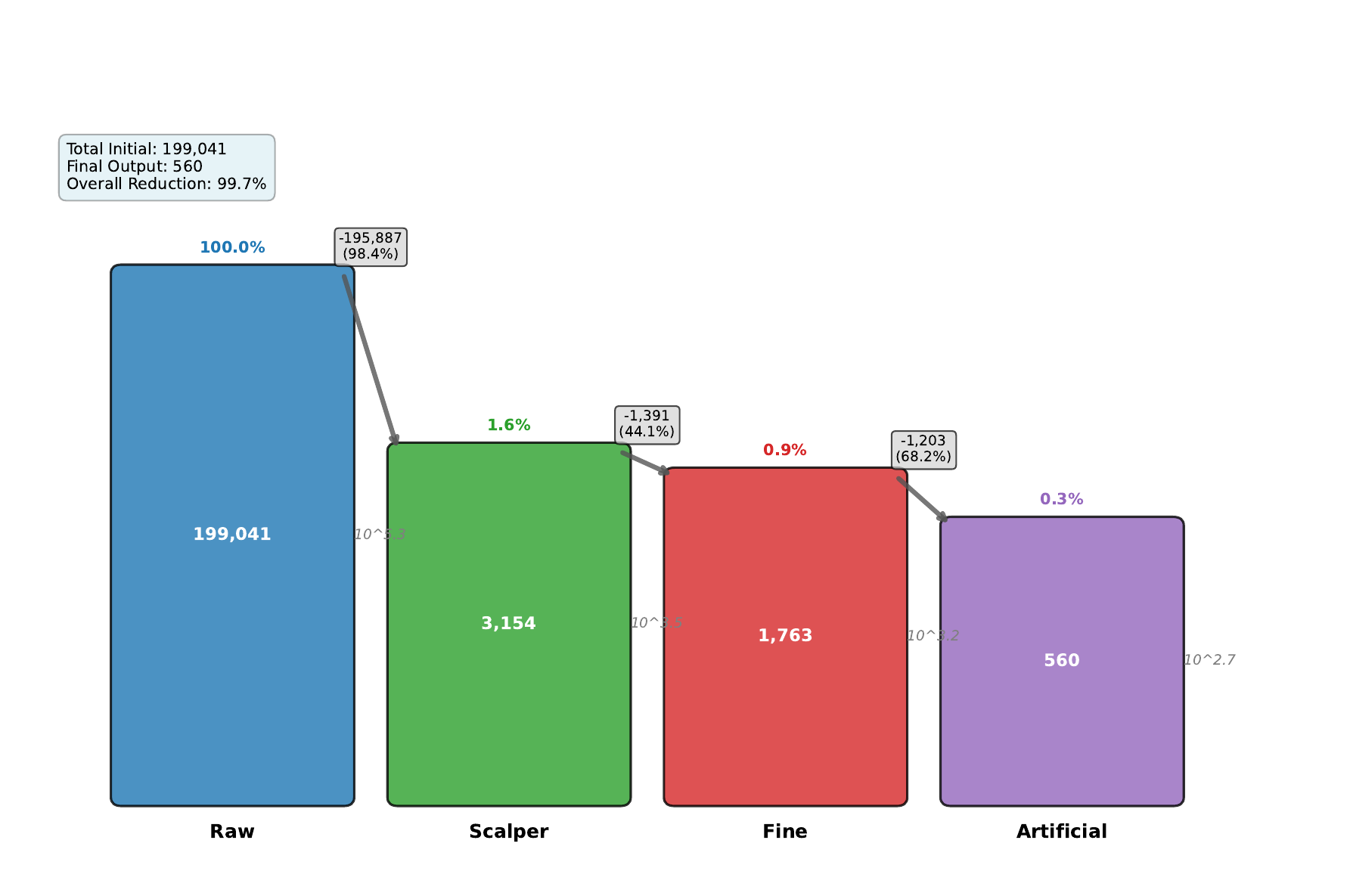}
  \caption{Document volume at each filtering stage of the benchmark construction pipeline, showing the reduction from raw retrieval results to the final final set.}
  \label{appendix:fig:benchmark-stage-volume}
\end{figure*}

\end{document}